\documentclass[11 pt]{article}
\usepackage{amssymb}
\usepackage{amsmath}
\usepackage{fullpage}
\usepackage[hmargin=1.0in,vmargin=1.0in]{geometry}

\newtheorem{Theorem}{\sc Theorem}

\newtheorem{Remark}{\sc Remark}

\newcommand{\BIGOP}[1]{\mathop{\mathchoice
{\raise-0.22em\hbox{\huge $#1$}}
{\raise-0.05em\hbox{\LARGE $#1$}}{\hbox{\large $#1$}}{#1}}}

\newenvironment{Proof}{\par  \sc
Proof.\rm}{\hspace*{\fill}$\triangle$\vspace{1ex}}

\def\BbbZ{{\sf Z\hspace*{-0.95ex}Z}}
\newcommand{\Z}[1]{\relax\ifmmode\BbbZ_{#1}\else $\BbbZ_{#1}$\fi}

\newcommand{\PP}{\mathbb P}

\newcommand{\ZZ}{\mathbb Z}

\newcommand{\F}{\mathbb F}

\newcommand{\cancel}[1]{}

\newcommand{\Ga}{\alpha}
\newcommand{\Gb}{\beta}
     
\newcommand{\Gd}{\delta}     
\newcommand{\Ge}{\epsilon}

\newcommand{\Gl}{\lambda}    
     
\newcommand{\Gs}{\sigma}

\newcommand{\mA}{\mathcal{A}}

\newcommand{\mL}{\mathcal{L}}

\newcommand{\mP}{\mathcal{P}}
\newcommand{\mQ}{\mathcal{Q}}

\newcommand{\ba}{{\bf a}}
\newcommand{\bb}{{\bf b}}
\newcommand{\bc}{{\bf c}}

\newcommand{\bu}{{\bf u}}
\newcommand{\bv}{{\bf v}}
\newcommand{\bw}{{\bf w}}
\newcommand{\bx}{{\bf x}}
\newcommand{\by}{{\bf y}}
\newcommand{\bz}{{\bf z}}

\def \ba {{\bf a}}
\def \bc {{\bf c}}
\def \bx {{\bf x}}

\def \by {{\bf y}}

\def \bz {{\bf z}}
\def \bu {{\bf u}}
\def \bv {{\bf v}}
\def\bw{{\bf w}}
\def \bo {{\bf 0}}

\def \wt {{\rm wt}}
\def\F{{\mathbb{F}}}

\def\proj{{\rm proj}}
\def\poly{{\rm Poly}}

\def\RM{{\rm RM}}

\def\BbbZ{{\sf Z\hspace*{-0.95ex}Z}}

\date{}

\def\xing#1 {\fbox {\footnote {\ }}\ \footnotetext { From Xing: {#1}}}
\def\hxing#1 {}

\newtheorem{theorem}{Theorem}[section]
\newtheorem{definition}[theorem]{Definition}
\newtheorem{proposition}[theorem]{Proposition}
\newtheorem{lemma}[theorem]{Lemma}
\newtheorem{corollary}[theorem]{Corollary}
\numberwithin{equation}{section} 
\newtheorem{example}[theorem]{Example}
\newtheorem{remark}[theorem]{Remark}

\begin{document}

\title{Efficient Multi-Point Local Decoding of Reed-Muller Codes via Interleaved Codex}
\author{Ronald Cramer\thanks{ CWI, Amsterdam and Mathematical Institute, Leiden University (email: Ronald.Cramer@cwi.nl) }, \; Chaoping Xing\thanks{ School of Physical and Mathematical Sciences, Nanyang Technological University, Singapore (email: xingcp@ntu.edu.sg)
}\; and\; Chen Yuan\thanks{ CWI, Amsterdam (email: chen.yuan@cwi.nl)
}}

\maketitle
\setcounter{page}{0}

\begin{abstract} Reed-Muller codes are among the most important classes of locally correctable codes. Currently  local decoding of Reed-Muller codes is  based on decoding on lines or quadratic curves to recover one single coordinate. To recover multiple coordinates simultaneously, the naive way is to repeat the local decoding for recovery of a single coordinate. This decoding algorithm might be more expensive, i.e., require higher query complexity. 

In this paper, we focus on Reed-Muller codes with usual parameter regime, namely, the total degree of evaluation polynomials is $d=\Theta({q})$, where $q$ is the code alphabet size (in fact, $d$ can be as big as $q/4$ in our setting).
By introducing  a novel variation of  codex, i.e., interleaved codex  (the concept of codex has been used for arithmetic secret sharing \cite{C11,CCX12}), we are able to locally recover arbitrarily large number $k$ of coordinates of a  Reed-Muller code simultaneously at the cost of querying $O(q^2k)$ coordinates. It turns out that our local decoding of Reed-Muller codes shows ({\it perhaps surprisingly}) that accessing $k$ locations is in fact cheaper than repeating the procedure for accessing a single location for $k$ times. Precisely speaking,
 to get the same success probability from repetition of local decoding for recovery of a single coordinate, one  has to query  $O(qk^2)$ coordinates. Thus, the query complexity of our local decoding is smaller for $k=\Omega(q)$. In addition, our local decoding is efficient, i.e., the decoding complexity is $\poly(k,q)$. Construction of an interleaved codex is based on concatenation of a codex with a multiplication friendly pair, while the main tool to realize codex is based on algebraic function fields (or more precisely, algebraic geometry codes).  Our estimation of success error probability is based on error probability bound for $t$-wise linearly independent variables given in \cite{BR94}.
\end{abstract}

\newpage

\section{Introduction}
In some applications such as transmission of information over noise channels or data storage, people are often interested in a portion of data. Thus, one needs to decode only this portion  of data instead of the whole data. However, classical error-correcting codes are generally used to recover the whole information. Thus, it is demanded to have a special class of error-correcting codes, i.e., locally decodable (correctable) codes.

Although locally decodable (correctable) codes have been studied for about two decades, Reed-Muller codes and their variants are still among the most important classes of locally correctable codes. Therefore, local decoding of Reed-Muller codes plays significant role in this topic. There are various decodings of Reed-Muller codes such as local decoding, list decoding or local list decoding in the literature \cite{AKKLR05,BL15,Go10,GJX15,PW04,SSV15}.  Among these decodings, there are basically two local decoding methods, i.e., decoding on lines and quadratic curves. Though decoding on quadratic curves can be generalized to decoding on higher power curves, it does not appear in the literature. Almost all locally correctable codes including Reed-Muller codes focus on correction of one single coordinate \cite{AKKLR05,BGKKS13,Go10,GJX15,HLW06,KSK14,Meir14,Ye12}.
To recover multiple coordinates simultaneously, the naive way is to repeat these local decodings of single coordinate. However, this idea does not work well when  locally recovering a large number of coordinates simultaneously is demanded (see Subsection 1.5 below).

 The current local decoding of Reed-Muller codes is based on decoding on lines or curves, i.e., randomly choose a line or a curve passing through the point where one intends to locally decode, then reduce it to the Reed-Solomon code decoding. Actually, in the PCP literature, one considers reading projection of a codeword to a low-degree curve instead of line~\cite{MR10}. However, the decoding algorithm is eventually reduced to decoding of Reed-Solomon codes again.  Therefore, for a fixed alphabet size, one could not read and decode coordinates as many as one wishes. Instead, one has to run decoding algorithm multiple times which increases error probability.

The main reason why the above local decoding of multiple points requires higher query complexity is that Reed-Solomon codes are used. Thus, it is nature to replace Reed-Solomon codes by algebraic geometry codes in local decoding for recovery of multiple coordinates. However, in order to apply algebraic geometry codes for local decoding of Reed-Muller codes, one has to consider certain t-wise independence to obtain good success probability from the Second t-wise Independence Tail Inequality. To achieve t-wise independence, we introduce a local decoding of Reed-Muller codes via a codex or a  variation of  codex, i.e., interleaved codex  (the concept of codex has been used for arithmetic secret sharing \cite{C11,CCX12}). It turns out that one can locally recover multiple coordinates of a Reed-Muller codeword simultaneously as long as there exists a good codex. On one hand, the only way to construct good codex is via algebraic curves over finite fields (or more precisely algebraic geometry codes). As algebraic function fields with many rational places are usually defined over $\F_{q^2}$, the codex built from these function fields are also defined over $\F_{q^2}$. Thus, we first need to reduce the field size from $q^2$ to $q$ to get an interleaved codex, and then locally decode Reed-Muller codes via interleaved codex.
The reduction technique is concatenation of codex over $\F_{q^2}$ with a multiplication friendly pair that was first introduced in \cite{DD87} to study multiplication of elements in extension fields of $\F_q$.
Essentially our local decoding of multiple coordinates is based on decoding of algebraic geometry codes which generalizes local decoding based on Reed-Solomon codes. However, this generalization is by no means trivial. In fact, several sophisticated algebraic tools are used to achieve our local decoding goal.

In this paper, we consider local decoding of  Reed-Muller codes with the usual parameter regime, i.e., $d=\Theta(q)$, where $q$ is the code alphabet size (in fact, $d$ can be as big as $q/4$ in our setting)
  As a main consequence of our local decoding, we are able to locally correct arbitrarily large number $k$ of coordinates simultaneously at the cost of querying $O(q^2k)$ coordinates. This is not achievable by all other existing local decodings of Reed-Muller codes. For instance, to get the same success probability from repetition of local decoding for recovery of a single coordinate, one  has to query  $O(qk^2)$ coordinates. Thus, the query complexity of our local decoding is smaller for $k=\Omega(q)$.
   Furthermore, our local decoding  is efficient, i.e., the decoding complexity is $\poly(k,q)$. In addition, our local decoding also works for recovery of one single coordinate as well. In this case, there is a trade-off between code dimension and success probability.

In the literature, there is a construction of locally decodable (correctable) codes via algebraic function fields (or algebraic curves) with large automorphism groups \cite{BGKKS13,Guo14}. However, usage of algebraic curves in the present paper is not for purpose of construction of locally correctable codes, but  local decoding of Reed-Muller codes.
\subsection{Locally correctable codes}
In order to state our result more accurately, let us introduce locally correctable codes first.

\begin{definition}\label{def:1.1}{\rm A subset $C$ of $\F_q^N$ is called a $q$-ary $(r,\Gd,\Ge)$-locally correctable code of length $N$ if there exists a randomized algorithm $\mA$ such that (i) for any $i\in[N]$ and $\bc\in C$, $\by\in\F_q^N$ with $\wt_H(\bc,\by)\le \Gd N$, one has $\Pr[\mA^{\by}(i)=c_i]\ge 1-\Ge$, where the probability is taken over random coin tosses of the algorithm $\mA$ (note that $c_i$ stands for the $i$-th coordinate of $\bc$ and $\mA^{\by}(i)$ stands for the output of $\mA$ from $\by$ for the position at $i$); (ii) $\mA$ makes at most $r$ queries to $\by$.
}\end{definition}

The above definition is only for recovery of one single coordinate (or point). We can generalize it to a locally correctable code with recovery of multiple coordinates (or points).

\begin{definition}\label{def:1.2}{\rm A subset $C$ of $\F_q^N$ is called a $q$-ary $(k; r,\Gd,\Ge)$-locally correctable code of length $N$ if there exists a randomized algorithm $\mA$ such that (i) for any $S\subseteq[N]$ with $|S|\le k$, and  $\bc\in C$, $\by\in\F_q^N$ with $\wt_H(\bc,\by)\le \Gd N$, one has $\Pr[\mA^{\by}(S)=\bc_S]\ge 1-\Ge$, where the probability is taken over random coin tosses of the algorithm $\mA$ (note that $\bc_S$ stands for the projection of $\bc$ to $S$  and $\mA^{\by}(S)$ stands for the output of $\mA$ from $\by$ for the positions at $S$); (ii) $\mA$ makes at most $r$ queries to $\by$.
}\end{definition}
Thus, a $(1; r,\Gd,\Ge)$-locally correctable code is  an $(r,\Gd,\Ge)$-locally correctable code.
\subsection{Reed-Muller codes}\label{subsec:RMcode}

We denote by $\bx$ the variable vector $(x_1,\dots,x_m)$. The multivariate polynomial ring $\F_q[x_1,\dots,x_m]$ is denoted by $\F_q[\bx]$.  For a vector $I=(e_1,\dots,e_m)\in \ZZ_{\ge 0}^m$, we denote by $\bx^I$ the monomial $\prod_{i=1}^mx_i^{e_i}$. Thus, we can write a polynomial of total degree at most $d$ by $f(\bx)=\sum_{\wt_L(I)\le d}a_I\bx^I$, where $a_I\in\F_q$ and $\wt_L(I)=\sum_{i=1}^me_i$ is the Lee weight. A polynomial in $\F_q[\bx]$ is called a degree-$d$ polynomial if its total degree is at most $d$. In the setting throughout the paper, we assume that $d < q$.
\begin{definition}\label{def:RM}{\rm The Reed-Muller code $\RM(q,d,m)$  is defined by $\{(f(\bu))_{\bu\in\F_q^m}:\; f(\bx)\in \F_q[\bx];$  $ \deg(f(\bx))\le d\}$, where $\deg(f(\bx))$ denotes the total degree of $f(\bx)$.
}\end{definition}
The dimension of the Reed-Muller code $\RM(q,d,m)$ is $\binom{m+d}{d}$.
 Currently, the two most popular parameter regimes for locally decoding Reed-Muller codes are either constant query complexity or $d\lesssim \sigma q$. In this paper, we focus on the case where $d\lesssim\Gs {q}$ for a fixed real $\Gs\in(0,1)$.

\subsection{Known results}
The simplest local decodings of Reed-Muller codes is called decoding on lines \cite[Propositions 2.5]{Ye12}. The  decoding on line can be generalized to decoding on quadratic curves \cite[Proposition 2.6]{Ye12}. Both these decodings are  very special cases of our codex decoding where a Reed-Solomon code with pairwise independent variables is used (see Example \ref{exm:4.1}(i) and (ii)).
\begin{proposition}\label{prop:1.2} Let $0<\Gs,\Gd<1$ be  positive real. Let $m$ and $d$ be positive integers. Let $q$ be a prime power.
 \begin{itemize}
 \item[{\rm (i)}] If $d \le\Gs(q-1)-1$, then the Reed-Muller code $\RM(q,d,m)$ is $(q-1,\Gd,2\Gd/(1-\Gs))$-locally correctable for all positive real with $\Gd<\frac{1-\Gs}2$.
     \item[{\rm (ii)}] If  $d \le\Gs(q-1)-1$, then the Reed-Muller code $\RM(q,d,m)$ is $\left(q-1,\Gd,\Ge=O\left(\frac{\gamma_{\Gs,\Gd}}{\sqrt{q}}\right)\right)$-locally correctable for all positive real with $\Gd<\frac{1-2\Gs}2$,   where $\gamma_{\Gs,\Gd}=\frac{\Gd-\Gd^2}{1-2\Gs-2\Gd}$.
\end{itemize}
\end{proposition}
The purpose of (ii) in Proposition \ref{prop:1.2} is to increase the success probability of local decoding. As $\Gs,\Gd$ are constant and $q$ is usually large, Proposition \ref{prop:1.2}(ii) gives much better success probability at the cost of a slightly smaller dimension.

Although it does not appear in the literature, generalization of local decoding on quadratic curves is quite straightforward in the following way. Assume that $f(\bu)_{\bu\in\F_q^m}$ is transmitted and we want to recover $f(\bw)$ at position $\bw$. Choose $t$ independently random vectors $\bv_1,\dots,\bv_t$ and consider the degree $t$ curve $\bw+\sum_{i=1}^tx^i\bv_i$. By using the error probability bound for $t$-wise independence (see Lemma \ref{twise}), we obtain the  result on local decoding using higher degree curves (see Example \ref{exm:4.1}(iii)).


\subsection{Our results}\label{subsec:1.4}
This paper mainly focuses on multiple point local decoding although single point local decoding is considered as well.

We consider local decoding of Reed-Muller codes via codex as well as interleaved codex.  If applying Reed-Solomon codes to our local decoding, we can use codex directly since we do not require that Reed-Solomon codes are defined over $\F_{q^2}$. However, if applying algebraic geometry codes from the Garcia-Stichtenoth tower, we have to get an interleaved codex over $\F_q$ from a codex over $\F_{q^2}$ and then do local decoding

For local decoding to recover multiple coordinates, we only state the result  based on the Garcia-Stichtenoth tower though all three classes of codes, namely Reed-Solomon codes, Hermitian codes and algebraic geometry codes from the Garcia-Stichtenoth tower are discussed in this paper.
We refer to Theorem \ref{thm:4.5}(i)-(iv) for local decoding of recovering multiple coordinates based on Reed-Solomon and Hermitian codes.

\begin{Theorem}\label{thm:2}  Let $q$ be a prime power. Let $d>1,  m, k$ be positive integers. Let $\Gd,\Gs$ be two reals in $(0,1)$ with $\Gd<\frac{1-4\Gs}2$ and $d<\Gs q$.
Then the Reed-Muller code $\RM(q,d,m)$   is   $\left(k; q^2k,\Gd,O\left(\left(\frac{\mu_{\Gd,\Gs}}{\sqrt{q}}\right)^k\right)\right)$-locally correctable, where $\mu_{\Gd,\Gs}=\frac{\sqrt{8}}{1-4\Gs-2\Gd}$ {\rm (}note that $k$ can be arbitrarily   large{\rm)}. Furthermore, the decoding algorithm is efficient, i.e., the decoding time complexity  is $\poly(k,q)$.
\end{Theorem}
\subsection{Comparison}
Let us compare our results given in Subsection \ref{subsec:1.4} with the known results (or those derived from the known results). 

\begin{itemize}
\item[(i)] To obtain a $k$-multiple point local decoding from the single point decoding given in Proposition \ref{prop:1.2}(ii), one can repeat local decoding $k$ times to get a $(k; qk, \Gd, \Ge)$-locally correctable code with $\Ge=O_{\Gs,\Gd}\left(\frac k{{q}}\right)$. Therefore, this method does not work when $k>{q}$.
           \item[(ii)] The other way is to first repeat local decoding to correct $f(\bu)$ at the same point $\bu$ to increase probability, and then repeat the above procedure to correct multiple points with meaningful probability. Let us analyze this decoding idea in detail.
           To increase decoding success probability of the local decoding in Proposition \ref{prop:1.2}(ii), we can repeat local correction of $f(\bu)$ at $\bu$ for $s$ times.
Denote by $Y_i$ a binary random variable such that $Y_i=1$ if the local decoding algorithm outputs a wrong answer in the $i$-th round and $Y_i=0$ otherwise. It follows from Proposition \ref{prop:1.2}(ii) that
$\Pr[X_i=1]=b= O\left(\frac{\gamma_{\Gs,\Gd}}{\sqrt{q}}\right)$. Thus, we have
\begin{equation}\label{eq:1.1}
\Pr\left[\sum_{i=1}^{s}Y_i\geq\frac{s}{2}\right]=\sum_{i\geq s/2}\binom{s}{i}b^i(1-b)^{s-i}=O\left(\left(\frac{4\gamma_{\Gs,\Gd}}{{q}}\right)^{s/2}\right).
\end{equation}
Therefore, we conclude that the Reed-Muller code $\RM(q,d,m)$ is $\left(qs,\Gd,\Ge'\right)$-locally correctable, where $\Ge'$ is given in \eqref{eq:1.1}. By repeating the above decoding procedure to correct $k$ points, we can also  conclude that the Reed-Muller code $\RM(q,d,m)$ is $\left(k;kqs,\Gd,k\Ge'\right)$-locally correctable.
\item[(iii)] By applying $k$-multiple point local decodings in Theorem \ref{thm:2}, the number $k$  is unbounded.  This means that we can recover any number $k$ of coordinates simultaneously with a high probability. At meanwhile, the number of queries is  $O(q^2k)$ (this is by no means possible for all other local decodings).
     \begin{itemize}
\item[(a)]
    By repeating the local decoding described in (ii), to correct $k$ points  with the same success probability $1-O\left(\left(\frac{\mu_{\Gs,\Gd}}{\sqrt{q}}\right)^k\right)$  as in Theorem \ref{thm:2}, $s$ in \eqref{eq:1.1} has to be $\Omega(k)$. Thus, the  decoding algorithm discussed in the above (ii) requires the query complexity  $\Omega(qk^2)$. This means that, for $k=\Omega(q)$, our local decoding of Reed-Muller codes in Theorem \ref{thm:2} is  cheaper than repeating the procedure for accessing a single location for $k$ times.
 \item[(b)]  Even for a unfair comparison, namely, in order to get a meaningful success probability $O(1)$ by repeating local decoding  of a single location for $k$ times, $s$ in \eqref{eq:1.1} has to be $\Omega(\log k/\log q)$. Thus, the  decoding algorithm discussed in the above (ii) requires the query complexity  $\Omega(qk\log k/\log q)$. In this case, for the parameter regime
    where the number $m$ of variables of evaluation polynomials is much bigger than the code alphabet size $q$, our local decoding of Reed-Muller codes in Theorem \ref{thm:2} is still cheaper than repeating the procedure for accessing a single location for $k$ times if $k=\Omega(q^q)$.
     \end{itemize}
\end{itemize}

\begin{Remark}{\rm One could consider the following local decoding. Assume that $f(\bu)_{\bu\in\F_q^m}$ is transmitted and we want to recover $f(\bw_i)$ at position $\bw_i$ for $i=1,2,\dots,k$.  Randomly choose $\bw\in\F_q^m$ and $\bu_1,\dots,\bu_m\in\F_q^e$ for some $e\ge k$ such that the plane $\bw+(\bu_1\cdot \by,\dots,\bu_m\cdot \by )$ passes through $\bw_1,\dots,\bw_k$, where $\by=(y_1,\dots,y_e)$ and  $\bu_i\cdot \by$ stands for the usual dot product.  Then  $f(\bw+(\bu_1\cdot \by,\dots,\bu_m\cdot \by ))$ is a polynomial of degree at most $d$. One can query at the point set $\{\bw+(\bu_1\cdot \bv,\dots,\bu_m\cdot \bv ):\; \bv\in\F_q^e\}$ to recover $f(\bw+(\bu_1\cdot \by,\dots,\bu_m\cdot \by ))$  as long as there are less than $(1-d/q)q^e/2$ error locations among these $q^e$ points. The query complexity of this local decoding is $q^e\ge q^k$ which is much bigger than $O(q^2k)$ for $k>q$.  We could replace linear polynomial vector by a lower degree polynomial vector $\bw+\sum_{j=1}^{\ell}(\bu_{1j}\cdot \by^j,\dots,\bu_{mj}\cdot \by^j )$  for local decoding, where $\by^j=(y_1^j,\dots,y_e^j)$. Then we have to require $\ell e\ge k$ and $\ell d<q$. As $q$ and $d$ are proportional, $\ell$ is a constant. In this case, the query complexity is still $q^e\ge q^{k/\ell}=q^{\Omega(k)}$.
}
\end{Remark}
\subsection{Our techniques}
Assume that $f(\bu)_{\bu\in\F_q^m}$ is transmitted for a degree-$d$ polynomial $f(\bx)$ and we want to recover $f(\bw)$ at position $\bw=(w_1,\dots,w_m)$. In the curve decoding, one replaces $(x_1,\dots,x_m)$ by $\bw+\Gl\bu_1+\Gl^2\bu_2$ for some random vectors $\bu_1=(u_{11},\dots,u_{1m}),\bu_2=(u_{21},\dots,u_{2m})\in\F_q^m$ (i.e., replace $x_i$ by $w_i+u_{1i}\Gl+u_{2i}\Gl^2$ for $i=1,2,\dots,m$). Then the function $f(\bw+\Gl\bu_1+\Gl^2\bu_2)$ becomes a univariate polynomial of degree at most $2d$. Thus, one can decode it via Reed-Solomon codes. A natural idea to generalize this decoding is to replace $x_i$ by $z_i$ for some function $z_i$ in some Riemann-Roch space $\mL(G)$ for an effective divisor $G$ of an algebraic curve with many rational points. Then $f(z_1,\dots,z_m)$ becomes a function in the Riemann-Roch space $\mL(dG)$ and thus one can recover the function $f(z_1,\dots,z_m)$ by using decoding of algebraic geometry codes.  If we want to  recover $f(\bw_i)$ for $i=1,2,\dots,k$, we can simply take some rational points $Q_1,\dots,Q_k$ on this curve such that $(z_1(Q_i),\dots,z_m(Q_i))$ are equal to $\bw_i$ for all $1\le i\le m$. Unlike the curve decoding using Reed-Solomon codes where independence is automatically satisfied due to a Vandermonde  matrix, here we have to consider independence of the functions $z_1,\dots,z_m$. We achieve this through the codex configuration introduced in \cite{C11,CCX12}.
 A codex is nicely implemented in our local decoding because of several properties of codex: (i) a codex has high randomness and uniformity; (ii) a codex provides independent variables that are needed in local decoding of Reed-Muller codes; (iii) a codex also allows correction of errors.

On the other hands, there are not many ways to construct codex. As far as we know, the only way to construct codex is through algebraic curves with many rational points (or more precisely algebraic geometry codes). We apply three classes of curves, i.e., projective line, Hermitian curve and the Garcia-Stichtenoth tower, to construction of codex and realize our local decoding. Since a good asymptotic tower is usually defined over $\F_{q^2}$, the codex built from such a tower is also defined over $\F_{q^2}$. Thus, we have to reduce the field size from $q^2$ to $q$. Our technique to achieve this reduction is concatenation of codex via multiplication friendly pairs. The multiplication friendly pairs that we employ are simply from Reed-Solomon codes.

 As for error probability, we make use of the error probability bound for $t$-wise linearly independent variables given in \cite{BR94}.

\subsection{Organization}
The paper is organized as follows. In Section 2, we introduce some preliminaries including definitions of codex and interleaved codex,  a construction of codex through algebraic geometry codes, construction of interleaved codex, error probability bounds and introduction to Hermitian curves the Garcia-Stichenoth tower. Our local decoding algorithms of Reed-Muller codes through codex and interleaved codex are presented in Section 3. Finally we apply various codex to decoding algorithms in Section 3 to obtain our main results in Section 4.

\section{Preliminaries}

\subsection{Codex}

The concept of codex was first introduced in \cite{C11,CCX12,CDN15} for the purpose of arithmetic secret sharing. A special case of  codex in this paper was implicitly introduced in \cite{CDM00,CC06}.

Let $\F_q$ be a finite field of $q$ elements. $\F_q^{*}$ denotes the multiplicative group of $\F_q$. Let $n,t,d,r$ be positive integers with $d\geq 2$ and $1\leq t<r\leq n$.
Vectors in the $\F_q$-vector space $\F_q^n$ are denoted in boldface. If ${\bf u}\in \F_q^n$, its coordinates are denoted as
$(u_i)_{i=1}^n$. Define ${\bf 1}=(1, \ldots, 1)\in \F_q^n$. The standard inner-product on $\F_q^n$ is denoted $\langle \cdot, \cdot \rangle$.
If $A\subset \{1, \ldots, n\}$ is  non-empty, $\pi_A$ denotes projection of $\F_q^n$ onto
the $A$-indexed coordinates, i.e.,  $\pi_A({\bf u})=(u_i)_{i\in A}$ for all ${\bf u}\in \F_q^n$.

\begin{definition}{\rm
For ${\bf u}, {\bf v}\in \F_q^n$,
 ${\bf u}*{\bf v}$ denotes the vector $(u_1v_1, \ldots, u_nv_n)\in \F_q^n$.
 For an $\F_q$-linear code $C\subset \F_q^n$, the $\F_q$-linear code $C^{*d}\subset \F_q^n$, the {\em $d$-th power of $C$},
is defined as the $\F_q$-linear subspace generated by all terms of the form ${\bf c}_1* \cdots * {\bf c}_d$ with
${\bf c}_1, \ldots, {\bf c}_d\in C$.
}\end{definition}

Note that if ${\bf 1}\in C$, then $C=C^{*1}\subset C^{*2}\subset \ldots \subset C^{*d}$.

\medskip
\noindent
Consider the following special case of an arithmetic secret sharing scheme (SSS for short)
which, in turn, is a special case of an arithmetic codex~\cite{CCX12}.

\begin{definition}\label{def:asss}{\rm
An $(n,t,d,r; \F_{q}^{k}/\F_q)$-codex is a pair $(C, \psi)$ such that the following conditions are satisfied:
\begin{itemize}
\item[{\rm (i)}]
$C\subset \F_q^n$ is an $\F_{q}$-linear code
and
$\psi: C \longrightarrow \F_{q}^{k}$ is a surjective $\F_q$-vector space morphism.

\item[{\rm (ii)}] It is {\em unital}, i.e., ${\bf 1}\in C$ and $\psi({\bf 1})={\bf 1}$.

\item[{\rm (iii)}] ({\em $t$-privacy with uniformity})
For each $A\subset \{1, \ldots, n\}$ with $|A|=t$,  the projection map
$$
\proj_{\psi, A}: C \longrightarrow \F_{q}^{k} \times \F_q^t,\qquad
{\bf c} \mapsto (\psi(\bc), \proj_A({\bf c}))
$$
is surjective, where $\proj_A({\bf c}$ is the projection of $\bc$ at $A$.
\item[{\rm (iv)}] ({\em $(d,r)$-product reconstruction})
The map $\psi$ extends uniquely to an $\F_q$-linear map
$\psi: C^{*d}\longrightarrow \F_{q}^{k}$
such that the following holds.
\begin{enumerate}
\item[{\rm (a)}]
$\psi$ satisfies the multiplicative relation
$$
\psi({\bf c}_1* \cdots * {\bf c}_d)= \psi({\bf c}_1)*\cdots *\psi({\bf c}_d)\in \F_q^k,
$$ for all ${\bf c}_1, \ldots, {\bf c}_d\in C$.
\item[{\rm (b)}] $C^{*d}$ has minimum distance at least $n-r+1$. Thus,
$\psi$ is $r$-wise determined, i.e.,
$
\psi({\bf z})={\bf 0},
$
for all ${\bf z} \in C^{*d}$ with $\proj_B({\bf z})={\bf 0}$
for some $B\subset \{1, \ldots, n\}$ with $|B|=r$.
\end{enumerate}
\end{itemize}
}\end{definition}

\begin{remark}{\rm \begin{itemize}\item[(i)]
{\em Uniqueness} of $\psi$ needs not be required separately, as it is implied by existence. Also note that, in fact,
$\psi({\bf c}_1* \cdots * {\bf c}_{d'})= \psi({\bf c}_1)*\cdots *\psi({\bf c}_{d'})$
for all ${\bf c}_1, \ldots, {\bf c}_{d'}\in C$ and all integers $d'$ with $1\leq d'\leq d$.
\item[(ii)] Given the above codex, we can define an arithmetic SSS, where each coordinate of $\bc$ is a share and $\psi(\bc)$ is the secret (please refer to \cite{CCX12} for the details).
    \end{itemize}
    }
\end{remark}

For the purpose of our local decoding, we have to introduce a variation of the above codex, i.e., interleaved codex.

\begin{definition}\label{def:2.4}{\rm
An $(n,\ell, t,d,r; \F_{q}^{k}/\F_q)$-interleaved codex is a pair $(C, \varphi)$ such that the following conditions are satisfied:
\begin{itemize}
\item[{\rm (i)}]
$C\subset \F_q^{n\ell}$ is an $\F_{q}$-linear code
and
$\varphi: C \longrightarrow \F_{q}^{k}$ is a surjective $\F_q$-vector space morphism.

\item[{\rm (ii)}] It is {\em unital}, i.e., ${\bf 1}\in C$ and $\varphi({\bf 1})={\bf 1}$.

\item[{\rm (iii)}] ({\em weak $t$-privacy with uniformity}) Let codewords of $C$ be indexed by  pairs $(i,j)\in[n]\times[\ell]$, i.e., every codeword is written as $(c_{ij})_{1\le i\le n;1\le j\le \ell}$. Then for each $1\le j\le \ell$ and    each $A\subset \{(1,j), \ldots, (n,j)\}$ with $|A|=t$,  the projection map
$$
\proj_{\varphi, A}: C \longrightarrow \F_{q}^{k} \times \F_q^t,\qquad
{\bf c} \mapsto (\varphi(\bc), \proj_A({\bf c}))
$$
is surjective.
\item[{\rm (iv)}] ({\em $(d,r)$-product reconstruction})
The map $\varphi$ extends uniquely to an $\F_q$-linear map
$\varphi: C^{*d}\longrightarrow \F_{q}^{k}$
such that the following holds.
\begin{enumerate}
\item[{\rm (a)}]
$\varphi$ satisfies the multiplicative relation
$$
\varphi({\bf c}_1* \cdots * {\bf c}_d)= \varphi({\bf c}_1)*\cdots *\varphi({\bf c}_d)\in \F_q^k,
$$ for all ${\bf c}_1, \ldots, {\bf c}_d\in C$.
\item[{\rm (b)}] $C^{*d}$ has minimum distance at least $n-r+1$. Thus,
$\varphi$ is $r$-wise determined, i.e.,
$
\varphi({\bf z})={\bf 0},
$
for all ${\bf z} \in C^{*d}$ with $\proj_B({\bf z})={\bf 0}$
for some $B\subset [n]\times[\ell]$ with $|B|=r$.
\end{enumerate}
\end{itemize}
}\end{definition}




\subsection{A construction of codex}
As far as we know, the only way to construct codex with $t=\Omega(n)$ is through algebraic geometry codes. In this subsection, we briefly introduce algebraic geometry codes and show how to construct codex.

 For the convenience of reader, we start with some definitions and notations. The reader may refer to \cite{Stich09,TV91}.

 An {\em algebraic function field}  over
$\F_q$ in one variable is a field extension $F \supset \F_q$ such
that $F$ is a finite algebraic extension of  $\F_q(x)$ for some
$x\in F$ that is transcendental over $\F_q$. It is assumed that
$\F_q$ is its full field of constants, i.e., the algebraic closure
of $\F_q$ in $F$ is $\F_q$ itself.

Let $\PP_F$ denote the set of places of $F$. A divisor is a formal sum $G=\sum_{P\in\PP_F}a_PP$, where $a_P$ are integers and are equal to zero except for finitely many $P$.
For a divisor $G$ of $F$, we define the Riemann-Roch space by
$\mL(G):=\{f\in F^*:\; {\rm div}(f)+G\ge 0\}\cup\{0\}.$
Then $\mL(G)$ is a finite dimensional space over $\F_q$ and its
dimension $\dim_{\F_q}(G)$ is determined by the Riemann-Roch theorem which
gives
\[\dim_{\F_q}(G)=\deg(G)+1-g(F)+\ell(K-G),\]
where $K$ is a canonical divisor of degree $2g(F)-2$, and $g(F)$ is the genus of $F$. Therefore, we
always have that $\dim_{\F_q}(G)\ge \deg(G)+1-g(F)$ and the quality holds
if $\deg(G)\ge 2g(F)-1$.

Let $k,t,n$ be positive integers. Suppose $Q_1,\ldots,Q_k,P_1\ldots,P_n$ are distinct rational places of a function field $F$ and denote by $\mQ$ and $\mP$ the set $\{Q_1,\dots,Q_k\}$ and $\{P_1,\dots,P_n\}$, respectively. Let $G$ be a divisor of $F$ such that ${\rm Supp}(G)\cap (\mP\cup\mQ)=\emptyset$. We define an algebraic geometry code of length $k+n$ as follows
$$
C(G;\mQ+\mP)=\left\{(f(Q_1),\ldots,f(Q_k),f(P_1),\ldots,f(P_n):f\in \mL(G))\right\}\subseteq\F_q^k\times\F_q^n.
$$
We also denote by $C(G;\mP)$ the code obtained from $C(G;\mQ+\mP)$ by puncturing the first $k$ positions.
\begin{proposition}\label{prop:2.6} Let $F$ be a function field of genus $g(F)$ with two disjoint sets $\mQ=\{Q_1,\dots,Q_k\}$ and $\mP=\{P_1,\dots,P_n\}$ of rational places.
Let $t\geq 1$, $d\geq 2, r\ge 1$ satisfy $n\geq r>d(2g(F)+k+t-1)$. For a positive divisor $G$ with $\deg(G)=2g(F)+k+t-1$ and ${\rm Supp}(G)\cap (\mP\cup\mQ)=\emptyset$,
let $C$ be the code $C(G;\mP)$ and define the map $\psi$ from $C$ to $\F_q^k$ given by $(f(P_1),\dots,f(P_n))\mapsto(f(Q_1),\dots,f(Q_k))$ {\rm (}note that the function $f$ is uniquely determine by $(f(P_1),\dots,f(P_n))${\rm )}.
Then $(C,\psi)$ is an $(n,t,d,r;\F_q^k/ \F_q)$-codex.
\end{proposition}

\begin{Proof} It is clear that $\psi$ is $\F_q$-linear and unital. To prove that $\psi$ is subjective, we consider the kernel of $\psi$. The kernel clearly has dimension $\dim_{\F_q}(G-\sum_{i=1}^kQ_i)$ which is equal to $\deg(G)-k-g(F)+1$ by the Riemann-Roch Theorem. Thus, the image of $\psi$ has dimension $\dim_{\F_q}(G)-(\deg(G)-k-g(F)+1)=k$. This implies that $\psi$ is surjective. As $\deg(G)-(t+k)= 2g(F)-1$, one can show $t$-privacy with uniformity in the same way.

Finally, we verify that it is $(d,r)$-product reconstruction. For a function $f\in \mL(G)\subseteq F$, we denote by $\bb_f$ and $\bc_f$ the words $(f(Q_1),\dots,f(Q_k))$ and $(f(P_1),\dots,f(P_n))$, respectively. Thus, one has $\psi(\bc_f)=\bb_f$ for any $f\in\mL(G)$. Furthermore, for $d$ codewords $\bc_{f_1}*\cdots*\bc_{f_{d}}$ in $C(G,\mP)$ we have $\psi(\bc_{f_1}*\cdots*\bc_{f_{d}})=\psi(\bc_{f_1\cdots f_{d}})=\bb_{f_1\cdots f_{d}}=\bb_{f_1}*\cdots*\bb_{f_{d}}=\psi(\bc_{f_1})*\cdots*\psi(\bc_{f_{d}})$. Now for $\bz\in C^{*d}$, we have $\bz\in C(d G,\mP)$. Thus, there exists a function $h\in\mL(d G)$ such that $\bz=\bc_h$. If $\pi_B(\bz)=0$, i.e., $h\in \mL(d G-\sum_{i\in B}P_i)$, then we must have $h=0$ since $d\deg(G)<r=|B|$. Hence, $\psi(\bz)=\bo$.

This completes the proof.
\end{Proof}

\begin{example}{\rm Consider the rational function field $F=\F_q(x)$, then $g(F)=0$. Let $\mQ$ and $\mP$ be the set $\{0\}$ and $\F_q\setminus\{0\}$. In this case, $k=1$ and $n=q-1$.
\begin{itemize}
\item[(i)] Choose $t=1$, then for any $1<d<r\le q-1$, there exists is a $(q-1,1,d,r;\F_q/ \F_q)$-codex.
\item[(ii)] Choose $t=2$, then for any $1<2d<r\le q-1$, there exists is a $(q-1,2,d,r;\F_q/ \F_q)$-codex.
\end{itemize}
}
\end{example}

\subsection{Concatenation of codex}\label{subsec:2.3}
As algebraic function fields with many rational places are usually defined over $\F_{q^2}$, the codex constructed from function fields in the previous subsection is defined over $\F_{q^2}$ as well. Thus, we have to reduce the field size form $q^2$ to $q$ through concatenation. In order to concatenate codex over $\F_{q^2}$, we need to introduce the  following multiplication friendly pair. Multiplication friendly pairs were first introduced by  D.V. Chudnovsky and G.V. Chudnovsky \cite{DD87} as
bilinear multiplication algorithms to study multiplication complexity in extension fields. In fact, a multiplication friendly pair is a special codex. The reader may refer to \cite{CDN15} for the detail.

\begin{definition}\label{def:5} A pair $(\pi, \phi)$ is called a $(d,k,m)_q$-multiplication friendly pair if $\pi$ is an $\F_q$-linear map from $\F_{q^k}$ to $\F_q^m$ and $\phi$ is   an $\F_q$-linear map from $\F_q^m$  to $\F_{q^k}$ such that $\pi(1)=(1,\dots,1)$ and $\phi(\pi(\Ga_1)\ast\cdots\ast\pi(\Ga_d))=\Ga_1\cdots\Ga_d$ for all $\Ga_i\in\F_q$. A $(2,k,m)_q$-multiplication friendly pair is also called a bilinear multiplication friendly pair.
\end{definition}

It is well known that, for a multiplication friendly pair $(\pi, \phi)$, the map $\pi$ is injective (see \cite[Lemma 3.1]{GX14} for instance). Furthermore, by using Reed-Solomon codes, one can construct the following   multiplication friendly pair (see \cite[Lemma 3,2 and Example 3.3]{GX14}).

\begin{lemma}\label{lem:2.9}
 If $k\ge 2$ and $q> d(k-1)$, then there exists a $(d,k,q)_q$-multiplication friendly pair $(\pi,\phi)$ such that $(\pi(\F_{q^k}))^{\ast d}$ is a $q$-ary linear code of length $m$ and relative minimum distance at least $1-d(k-1)/q$.
\end{lemma}

Now, we proceed to concatenate a codex over $\F_{q^2}$ with a multiplication friendly pair given in Lemma \ref{lem:2.9}.

\begin{proposition}\label{prop:interleave} Given an $(n, t,d,r; \F_{q^2}^{k}/\F_{q^2})$-codex, one can construct an $(n, q, t,d,dn+qr; \F_{q^2}^{k}/\F_{q^2})$-interleaved codex in time $\poly(n,q)$.
\end{proposition}
\begin{Proof} Let $(C,\psi)$ be an $(n,t,d,r; \F_{q^2}^{k}/\F_{q^2})$-codex. By Lemma \ref{lem:2.9}, we have a  $(d,2,q)_q$-multiplication friendly pair $(\pi,\phi)$. We extend $\pi$ to an $\F_q$-linear map from $\F_{q^2}^{s}$ to $ \F_q^{qs}$ by defining $\pi(v_1,\ldots,v_{s})=(\pi(v_1),\dots,\pi(v_{s}))$ for every $s\ge 1$. Then it is clear that $\pi$ is injective on $\F_{q^2}^{s}$.

Put $C_1=\pi(\psi^{-1}(\F_q^k))\subseteq \F_q^{qn}$. Then $\pi^{-1}(C_1)=\psi^{-1}(\F_q^k)$ since $\pi$ is injective. For a codeword $\bc=(c_1,\dots,c_n)\in C$, we denote $\pi(c_i)$ by $(c_{i,1},\dots,c_{i,q})$. Thus, a codeword $\pi(\bc)$ of $C_1$ has coordinates indexed by pairs $(i,j)\in[n]\times[q]$.

Consider the maps
\[C_1=\pi(\psi^{-1}(\F_q^k))\xrightarrow{\pi^{-1}} \psi^{-1}(\F_q^k)\xrightarrow{\psi}\F_q^k.\]
Let $\varphi$ be the composition map $\psi\circ\pi^{-1}$. Then it is clear that $\varphi$ is an $\F_q$-linear map from $C_1$ to $\F_q^k$.  We claim that the pair $(C_1,\varphi)$ is the interleaved codex with desired parameters.

It is clear that $\varphi$ is  surjective.

As $\pi$ maps $1$ to the all-one vector of length $q$ and the all-one vector of length $n$ belongs to $C$, we conclude that the all-one vector ${\bf 1}$ of length $qn$ belongs to $C_1$. From the definition of $\varphi$, we clearly have $\varphi({\bf 1})={\bf 1}$.

To show $t$-weak privacy, let $(\bu,\bv)$ be a vector of $\F_{q}^{k} \times \F_q^t$. Then there is a vector $\bv'\in\F_{q^2}^t$ such that $\proj_j\circ\pi(\bv')=\bv$, where $\proj_j$ is the projection map of $\F_q^q$ at position $j$. Let $\bb\in C$ such that $(\psi(\bb),\proj_B(\bb))=(\bu,\bv')$ with $B=\{1\le i\le n:\; (i,j)\in A\}$.  Then $\bb$ belongs to $\psi^{-1}(\F_q^k)$.
 Now it is easy to verify that
$(\varphi(\bc), \proj_A({\bf c}))=(\bu,\bv)$, where $\bc=\pi(\bb)$.

Now, we move to  proof of the multiplication property. Note that $\phi$ is equal to $\pi^{-1}$ when restricted to $\pi(\F_{q^2})$. Thus, we can extend $\pi^{-1}$ to a map from  $\F_q^{qn}$ to $\F_{q^2}^n$ via replacement of  $\pi^{-1}$ by $\phi$. Thus, $\varphi$ is equal to $\psi\circ\phi$ on $C_1$. Hence,  $\varphi$  can be extended to a map from $C_1^{*d}$ to $\F_q^k$.

For $d$ vectors $\pi(\bc_1),\ldots,\pi(\bc_d)\in C_1$ with $\bc_i\in \psi^{-1}(\F_q^k)\subset C$,
we have
\begin{eqnarray*}
\varphi(\pi(\bc_1)\ast \cdots\ast\pi(\bc_d))&=&(\psi\circ\phi)(\pi(\bc_1)\ast \cdots\ast\pi(\bc_d))\\
&=&\psi(\bc_1\ast \cdots\ast\bc_d)=\psi(\bc_1)\ast \cdots\ast\psi(\bc_d)\\
&=&(\psi\circ\phi(\pi(\bc_1)))\ast \cdots\ast(\psi\circ\phi(\pi(\bc_d)))=
\varphi(\pi(\bc_1))\ast\cdots\ast\varphi(\pi(\bc_d)).
\end{eqnarray*}

Finally, note that $C_1$ is the concatenated code of $C$ with a $[q,2,q-1]$-Reed-Solomon code. Since $C^{*d}$ has minimum distance at least $n-r+1$ and $\pi(\F_{q^2})^{*d}$ has minimum distance at least $q-d$, we conclude that the minimum distance of $C_1^{*d}$ is at least $(n-r+1)(q-d)\ge qn-(dn+qr)+1$.
The proof is completed.
\end{Proof}

\subsection{A property of  codex}\label{subsec:RM}
Let $(C,\psi)$ be an $(n,t,d,r,\F_{q^2}^k/\F_{q^2})$-codex. Let $(C_1,\varphi)$ be the interleaved codex constructed from $(C,\psi)$  in {Proposition} \ref{prop:interleave}. Let $m$ be a positive integer.  For each integer $e\ge 1$ and each polynomial $f(\bx)\in \F_q[x_1, \ldots, x_m]$ with $\deg(f(\bx)) \leq d$.
Define the map
$
f^{(e)}: \F_q^{e \times m}\longrightarrow \F_q^e;\
({\bf u}_1, \ldots, {\bf u}_m)\mapsto (f(u_{1j}, \ldots, u_{mj}))_{i=1}^e,
$
where $u_{ij}$ denotes the $j$-th coordinate of ${\bf u}_i$ ($i=1, \ldots, m$, $j=1,\ldots, r$). Note that $f(u_1,\dots,u_m)=f^{(1)}(u_1,\dots,u_m)$.

For codewords $\bc_1,\dots,\bc_m\in C\subseteq\F_q^n$, we have
\begin{equation}f^{(n)}(\bc_1,\dots,\bc_m)=(f(\bc_{(1)}),\dots,f(\bc_{(m)})=(\cdots,\sum_{\wt_L(I)\le d}a_I\bc_{(j)}^{I},\cdots)=\sum_{\wt_L(I)\le d}a_I(\cdots,\bc_{(j)}^{I},\cdots),\end{equation} where $\bc_{(j)}^{I}=\prod_{i=1}^m{c_{ij}}^{e_i}$ for $I=(e_1,e_2,\dots,e_m)$. This implies that $f^{(n)}({\bf c}_1, \ldots, {\bf c}_m)\in C^{*d}$.
 Furthermore, we have
\[
\psi(f^{(n)}({\bf c}_1, \ldots, {\bf c}_m))=\sum_{\wt_L(I)\le d}a_I\psi(\cdots,\bc_{(j)}^{I},\cdots)=f^{(k)}(\psi(\bc_1),\dots,\psi(\bc_m))\]
and
\[
\varphi(f^{(n)}({\bf c}_1, \ldots, {\bf c}_m))=\sum_{\wt_L(I)\le d}a_I\varphi(\cdots,\bc_{(j)}^{I},\cdots)=f^{(k)}(\varphi(\bc_1),\dots,\varphi(\bc_m)).\]

\subsection{Bounds on error probability}

In this subsection, we study sum of $t$-wise independent variables that will be used in local decoding of Reed-Muller codes. For our purpose, let us consider binary random variables that take values either $0$ or $1$.
\begin{definition}\label{def:independ}{\rm  Binary random variables $X_1,X_2,\dots,X_n$ are said {\it $t$-wise independent} if for any $a_1,a_2,\dots,a_t\in\{0,1\}$ and any $t$ indices $1\le i_1<i_2<\cdots<i_t\le n$, one has $\Pr[X_{i_1}=a_1,\dots, X_{i_t}=a_t]=\prod_{i=1}^t\Pr[X_{i_i}=a_i]$.
}\end{definition}

We are going to bound the deviation from the mean of the sum $X=X_1+\cdots+X_n$. Let us first consider the case $t=2$ where Chebyshev's inequality is employed.

\begin{lemma}\label{lemma:chebyshev}
Let $X_1,\ldots,X_n$ be pairwise independent binary random variables taking values in $\{0,1\}$ and satisfy $\Pr(X_i=1)=\Gd$ for all $1\le i\le n$. Then, for any $A>0$, $\Pr[|X-\Gd n|\geq A]\leq  \frac{(\delta-\delta^2)n}{A^2}$.
\end{lemma}
\begin{Proof}
 Define $X=\sum_{i=1}^n X_i$.
By linearity of expectation,
$
\mathrm{E}[X]=\sum_{i=1}^n \mathrm{E}[X_i]=\delta n.
$
Since the $X_i$'s are pairwise independent, linearity of variance holds here as well.
This implies
$$
\mathrm{Var}(X)=\sum_{i=1}^M \mathrm{Var}[X_i]= \sum_{i=1}^n (\mathrm{E}[X_i^2]-\mathrm{E}[X_i]^2)=(\delta-\delta^2)n.
$$
Then by Chebyshev's Inequality, we have

$$
 \mathrm{Prob}[|X-\mathrm{E}[X]|\geq A]
  \leq    \frac{\mathrm{Var}(X)}{A^2}=
 \frac{(\delta-\delta^2)n}{A^2}.
$$
This completes the proof.
\end{Proof}

For $t\ge 4$, we have the following {\it Second $t$-wise Independence Tail Inequality} .
\begin{lemma}\label{twise} {\rm (see \cite{BR94})} Let $t\geq 4$ be an even integer. Suppose $X_1,\ldots,X_n$ are $t$-wise independent random variables over $\{0,1\}$. Let
$X:=\sum_{i=1}^{n}X_i$ and define $\mu:=E[X]$ be the expectation of the sum. Then, for any $A>0$, $Pr[|X-\mu|\geq A]\leq 8\left(\frac{t\mu+t^2}{A^2}\right)^{t/2}$.
\end{lemma}

\subsection{Two classes of function fields}\label{subsec:curve}
In this subsection, we introduce two classes of algebraic curves (or equivalently function fields) that will be used to construct our codex in Section \ref{sec:decoding}, namely Hamitian curves and the Garcia-Stichtenoth tower. The reader may refer to \cite{GS95} and \cite[Sections 6.4 and 7.2]{Stich09} for the details.

For a function $F$ of genus $g(F)$ over $\F_{q^2}$, the number $N(F)$ of rational places of $F$ is upper bounded by the Hasse-Weil bound $q+1+2g(F){q}$. $F$ is called maximal if $N(F)$ achieves the Hasse-Weil bound, i.e., $N(F)=q+1+2g(F){q}$. One of maximal function fields is called the Hermitian function field. It is defined over $\F_{q^2}$  and its equation is given by
\[y^q+y=x^{q+1}.\]
The function field of this curve is $F=\F_{q^2}(x,y)$. There are totally $q^3+1$ rational places for this function field. One of them is the point ``at infinity", denoted by $\infty$. The other places are  given by $(\Ga,\Gb)\in\F_{q^2}^2$ satisfying $\Gb^q+\Gb=\Gb^{q+1}$. These are called ``finite" rational places. The genus of this function field is $g(F)=q(q-1)/2$.

The other class of function fields is also defined over $\F_{q^2}$. It is asymptotically optimal and recursively defined by the following equations
\[x_{i+1}^q+x_{i+1}=\frac{x_i^q}{1+x_i^{q-1}},\quad i=1,2\dots\]
with $x_1$ being a transcendental element over $\F_q$. The  function field $\F_q(x_1,x_2,\dots,x_e)$ is denoted by $F_e$. The genus $g_e:=g(F_e)$ is at most $q^e$. There is one place over the pole of $x_1$ called ``point at infinity". Furthermore, for each element $\Ga\in\F_{q^2}\setminus\{\Ga\in\F_{q^2}:\; \Ga^q+\Ga=0\}$, there are exactly $q^{e-1}$ places over it. Thus,  the number $N(F_e)$ of rational places of $F_e$ is at least $q^e(q-1)+1$. Thus, one has
$\lim_{e\rightarrow\infty}N(F_e)/g(F_e)\ge q-1$. By the Vl\u{a}du\c{t}-Drinfeld bound \cite{TV91}. We must have $\lim_{e\rightarrow\infty}N(F_e)/g(F_e)={q}-1$.

\section{Local Decoding of Reed-Muller Codes}\label{sec:decoding}

In this section, we analyze local decoding of Reed-Muller codes to recover multiple coordinates simultaneously.
Let $\RM(q,d,m)$  be the $q$-ary Reed-Muller code. We denote by $\ba_f$ the codeword of  $\RM(q,d,m)$ generated by the polynomial $f(\bx)$. Let $N=q^m$ and $\Gd\in(0,1)$. Suppose $\ba_f$ is transmitted and there are at most $\Gd N$ error  positions, i.e., there exists a vector ${\bf b}\in \F_q^{N}$ with
$
\wt_{\mathrm{H}}({\bf b})\leq \delta N
$
such that the received word is
$
\tilde{{\bf a}}:= {\bf a}_f + {\bf b}\in \F_q^{N}.
$

In other words, $\tilde{{\bf a}}$ is a corruption of the codeword ${\bf a}_f$ by an error vector
${\bf b}$ of relative Hamming weight at most $\delta$. Assume that we are going to recover ${\bf a}_f$ at positions $\bw_1,\bw_2,\dots,\bw_{k}\in\F_q^m$. Write $\tilde{{\bf a}}=(\tilde{a}_{\bu})_{\bu\in\F_q^m}$ and $\bw_i=(w_{i,1},w_{i,2},\dots,w_{i,m})$ for $i=1,2,\dots,k$.
\subsection{Direct decoding with codex} We first introduce a local decoding with codex  without concatenation.\\ \medskip

\fbox{\begin{minipage}{35em}
\begin{center} {\bf Algorithm 1: Local Decoding Algorithm with Codex }\end{center}
\begin{itemize}
\item[1.] Choose  an  $(n,t,d,\Gs n,\F_q^{k}/\F_q)$-codex ${\cal C}=(C,\psi)$ with a real $0<\Gs<1$;
\item[2.]
For $i=1, \ldots, m$, select ${\bf c}_i\in C\subset \F_{q}^n$ uniformly at random (and independently of everything else) such that
$
\psi({\bf c}_i)=(w_{1,i},\dots,w_{k,i});
$
\item[3.] Query $\tilde{{\bf a}}=(\tilde{a}_{\bu})_{\bu\in\F_q^m}$ at positions $\bv_1,\bv_2,\dots,\bv_n\in\F_q^m$, where ${\bf v}_j$ denotes collection of the $j$-th coordinate of the codewords ${\bf c}_1, \ldots, {\bf c}_m$;
\item[4.] Find a codeword $(z_1,z_2,\dots,z_n)\in C^{*d}$ such that the Hamming distance between $(z_1,z_2,\dots,z_n)\in C^{*d}$  and $(\tilde{a}_{\bv_1},\dots,\tilde{a}_{\bv_n})$ is at most $(n-\Gs n)/2$.
\item[5.] If no such a codeword $(z_1,z_2,\dots,z_n)$ in Step 4 is found, output ``fail". Otherwise, output $(f(\bw_1),f(\bw_2),\dots,f(\bw_{k}))=\psi(z_1,z_2,\dots,z_n)$.
\end{itemize}
\end{minipage}}

\bigskip

Now, we analyze the above algorithm.

First, ${\bf v}_1, \ldots, {\bf v}_n$ are $t$-wise independent and uniformly random distributed in $\F_q^m$
 by Definition \ref{def:asss}(iii).

Suppose that a codeword $(z_1,z_2,\dots,z_n)\in C^{(d)}$ is found such that the Hamming distance between $(z_1,z_2,\dots,z_n)\in C^{*d}$  and $(\tilde{a}_{\bv_1},\dots,\tilde{a}_{\bv_n})$ is at most $(n-\Gs n)/2$. Then by Definition \ref{def:asss}(iv)(b), we have
$\psi(f({\bf v}_1), \ldots, f({\bf v}_n))=\psi(z_1,z_2,\dots,z_n) $ as long as the Hamming distance between $(f({\bf v}_1), \ldots, f({\bf v}_n))$ and $(\tilde{a}_{\bv_1},\dots,\tilde{a}_{\bv_n})$ is at most $(n-\Gs n)/2$.

 \noindent
By Subsection \ref{subsec:RM}, it holds that
$
f^{(n)}(\bc_1,\dots,\bc_m)=(f({\bf v}_1), \ldots, f({\bf v}_n))\in C^{*d}
$ and \\ $f^{(k)}(\psi(\bc_1),\dots,\psi(\bc_{m}))=(f(\bw_1),\dots,f(\bw_{k}))$.
Thus, we can recover $(f(\bw_1),\dots,f(\bw_{k}))$ as follows.
\begin{eqnarray*}
(f(\bw_1),\dots,f(\bw_{k}))&=&f(\psi(\bc_1),\dots,\psi(\bc_m))
=\psi(
f^{(n)}(\bc_1,\dots,\bc_m))\\ &= &\psi(f({\bf v}_1), \ldots, f({\bf v}_n))=\psi(z_1,z_2,\dots,z_n).
\end{eqnarray*}
Now the probability of successfully recovering $(f(\bw_1),\dots,f(\bw_{k}))$ is equal to the probability of successfully finding a codeword $(z_1,z_2,\dots,z_n) $ such that the Hamming distance between $(\tilde{a}_{\bv_1},\dots,\tilde{a}_{\bv_n})$ and $(z_1,z_2,\dots,z_n)$ is at most $(n-\Gs n)/2$. This probability is at least the probability that  there are at most $(n-\Gs n)/2$ corrupted positions  for $\ba_f$ among $\bv_1,\bv_2,\dots,\bv_n$.

Denote by $E$ the set of coordinates $\bu$ such that $\bb_{\bu}\neq0$.
For $j=1, \ldots, n$,  define the binary random variable $X_j$ such that $X_j=1$ if
${\bf v}_j\in E$ and $X_j=0$ otherwise. Then $X_1, \ldots, X_n$ are $t$-wise independent and
$\mbox{Prob}(X_j=1)=\delta$ for $j=1, \ldots, n$. Put $X=\sum_{i=1}^nX_i$.

Since the minimum distance of $C^{*d}$ is at least $(n-\sigma n)+1$, one can correctly recover  $\psi(z_1,z_2,\dots,z_n)$ from $(\tilde{a}_{\bv_1},\dots,\tilde{a}_{\bv_n})$ if $|E\cap\{\bv_1,\dots,\bv_n\}|\le (n-\Gs n)/2$.

Thus, by the above identity it implies that one can correctly recover $(f(\bw_1),\dots,f(\bw_{k}))$ with probability at least $1-\Pr(X\le (n-\Gs n)/2)$ by
querying $\tilde{{\bf a}}=(\tilde{a}_{{\bf u}})_{{\bf u}\in \F_q^m}$, at coordinates ${\bf v}_1, \ldots, {\bf v}_n$.

Summarizing the above analysis, we get the following local decoding of Reed-Muller codes.
\begin{theorem}\label{thm:3.1}
If there exists  an  $(n,t,d,\Gs n,\F_q^{k}/\F_q)$-codex $(C,\psi)$ with a real $0<\Gs<1$, then the Reed-Muller code $\RM(q,d,m)$ is an
$(k; n,\Gd,\Ge)$-locally decodable code with $\Ge=\Pr(X>(n-\Gs n)/2)$, where $X$ is defined above. Furthermore, the local decoding complexity is $\poly(n,k,q)$ if the codex can be constructed in time $\poly(n,,k,q)$ and decoding time of the code $C^{*d}$ is $\poly(n,q)$.
\end{theorem}

\subsection{Decoding with interleaved codex }\label{subsec:3.2} Now we introduce a local decoding with interleaved codex. We start with a codex over $\F_{q^2}$ and assume that $d\le \Gs q$ with $\Gs<1$.\\ \medskip

\fbox{\begin{minipage}{35em}
\begin{center} {\bf Algorithm 2: Local Decoding Algorithm with Interleaved Codex}\end{center}
\begin{itemize}
\item[1.] Choose  an  $(n,t,d,\rho n,\F_{q^2}^{k}/\F_{q^2})$-codex ${\cal C}=(C,\psi)$ with a real $0<\rho<1-\Gs$ and let $(C_1,\varphi)$ be the interleaved codex constructed from $(C,\psi)$ in {Proposition} \ref{prop:interleave};
\item[2.]
For $i=1, \ldots, m$, select ${\bf c}_i\in C\subset \F_{q^2}^n$ uniformly at random (and independently of everything else) such that
$
\varphi(\pi({\bf c}_i))=(w_{1,i},\dots,w_{k,i});
$
\item[3.] Query $\tilde{{\bf a}}=(\tilde{a}_{\bu})_{\bu\in\F_q^m}$ at positions $\bv_1,\bv_2,\dots,\bv_{qn}\in\F_q^m$, where ${\bf v}_j$ denotes collection of the $j$-th coordinate of the codewords $\pi({\bf c}_1), \ldots, \pi({\bf c}_m)$;
\item[4.] Find a codeword $\bz\in C_1^{*d}$ such that the Hamming distance between $\bz$  and $(\tilde{a}_{\bv_1},\dots,\tilde{a}_{\bv_{qn}})$ is at most $(1-\Gs-\rho )qn/2$.
\item[5.] If no such a codeword $\bz$ in Step 4 is found, output ``fail". Otherwise, output $(f(\bw_1),f(\bw_2),\dots,f(\bw_{k}))=\varphi(\bz)$.
\end{itemize}
\end{minipage}}

\bigskip

Analysis of the above algorithm is similar to that of Algorithm 1. Let us discuss probability only.

Note that $C_1$ is a concatenated code. The outer code is $C$ which is defined over
$\F_{q^2}$ and the inner code is a Reed-Solomon code. Thus, $(\tilde{a}_{\bv_1},\dots,\tilde{a}_{\bv_{qn}})$ can be partitioned into $n$ blocks $(\tilde{\ba}_1,\dots,\tilde{\ba}_n)$, each with length $q$. Write  $\tilde{\ba}_i=(\tilde{a}_{i,1},\ldots,\tilde{a}_{i,q})$ be the $i$-th block.
 Denote by $X_{i,j}$ for $(i,j)\in [n]\times[q]$ be the random variable such that
$X_{i,j}=1$ if $\tilde{a}_{i,j}$ is corrupted,  and $X_{i,j}=0$ otherwise. Then $\Pr[X_{i,j}=1]=\delta$ follows from the fact that there is $\delta$
fraction of  corrupted positions.
By $t$-weak privacy of the pair $(C_1,\varphi)$, it is clear that the random variable
$X_{1,j},X_{2,j},\ldots,X_{n,j}$ is $t$-wise independent. Let $X_i=\sum_{j=1}^n X_{j,i}$.
In Lemma \ref{twise},  put $A=(1-\Gs-\rho)n/2-\Gd n$, we obtain
$$
\Pr\left[X_i>\frac{(1-\Gs-\rho) n}{2}\right]\leq 8\left(\frac{4t\delta n +4t^2}{(1-\Gs-\rho-2\delta)^2n^2}\right)^{t/2}.
$$
By the union bound, we have
$$
\Pr\left[\exists i: X_i>\frac{(1-\Gs-\rho)n}{2}\right]\leq 8q\left(\frac{4t\delta n +4t^2}{(1-\Gs-\rho-2\delta)^2n^2}\right)^{t/2}.
$$
Thus, we have
\begin{equation}\label{eq:a1}
\Pr\left[\sum_{i=1}^{q}X_i>\frac{(1-\sigma-\rho)qn}{2}\right]
\le \Pr\left[\exists i: X_i>\frac{(1-\sigma-\rho)n}{2}\right]  \\
\leq 8q\left(\frac{4t\delta n +4t^2}{(1-\Gs-\rho-2\delta)^2n^2}\right)^{t/2}.
\end{equation} $C_1^{*d}$ is a concatenated code and it has minimum distance  at least $qn-dn-\Gs qn+1=qn(1-\Gs-\rho)+1$. By \cite{F66}, we know that a concatenated code can be efficiently decoded up to half of minimum distance. This completes analysis of Algorithm 2.

Summarizing the above analysis, we get the following local decoding of Reed-Muller codes.
\begin{theorem}\label{thm:3.2} Let $d\le \Gs q$.
If there exists  an  $(n,t,d,\rho n,\F_{q^2}^{k}/\F_{q^2})$-codex $(C,\psi)$ with a real $0<\rho<1-\Gs$, then the Reed-Muller code $\RM(q,d,m)$ is a
$(k; n,\Gd,\Ge)$-locally decodable code with $\Ge$ upper bounded by \eqref{eq:a1}. Furthermore, the local decoding complexity is $\poly(n,k,q)$ if the codex can be constructed in time $\poly(n,k,q)$ and decoding time of the code $C_1^{*d}$ is $\poly(n,q)$, where $C_1$ is the concatenated code defined in Subsection \ref{subsec:2.3}.
\end{theorem}

\section{The main results}
In this section, we apply  various codex constructed from the rational function fields, Hermitian function fields and function fields in the Garcia-Stichtenoth tower to obtain our main results by using Theorems \ref{thm:3.1} or \ref{thm:3.2}.
\subsection{Single point decoding}
In this subsection, we consider local decoding to recover only a single coordinate via codex from Reed-Muller codes.
\begin{example}\label{exm:4.1}{\rm  For the rational function field $F=\F_q(x)$, we have $g(F)=0$. Let $\mQ$ and $\mP$ be the set $\{0\}$ and $\F_q\setminus\{0\}$. In this case, $k=1$ and $n=q-1$.
\begin{itemize}
\item[(i)] Choose $t=1$, then for any real $0<\Gs< 1$ and $1<d\le \Gs(q-1)+1$, there exists is a $(q-1,1,d,\Gs(q-1);\F_q/ \F_q)$-codex. By  Markov's inequality the probability that $(1-\Gs)(q-1)/2 $ or more of the queries go to corrupted locations is at most $2\Gd/(1-\Gs)$. Thus, the Reed-Muller code $\RM(q,d,m)$ is a
$(q-1,\Gd,2\Gd/(1-\Gs))$-locally correctable code by Theorem \ref{thm:3.1}. This is exactly the same decoding given in \cite[Proposition 2.5]{Ye12}.
\item[(ii)] Choose $t=2$, then for any real $0<\Gs< 1$ and $1<d\le \Gs(q-1)-1$, there exists is a $(q-1,2,d,2\Gs(q-1);\F_q/ \F_q)$-codex. In Lemma \ref{lemma:chebyshev}, let $A$ be $(1-2\Gs)(q-1)/2-\Gd (q-1)$, we obtain
\begin{equation}\label{eq:ge}\Ge=\Pr[X>(1-2\Gs )(q-1)/2]\le\frac{(\Gd-\Gd^2)(q-1)}{((1-2\Gs)(q-1)/2-\Gd (q-1))^2}=\frac{4(\Gd-\Gd^2)}{(1-2\Gs-2\Gd)^2}\times\frac1{q-1}.\end{equation}
Thus,  by Theorem \ref{thm:3.1}, the Reed-Muller code $\RM(q,d,m)$ is a
$(q-1,\Gd,\Ge)$-locally correctable code with $\Ge$ given in \eqref{eq:ge}.
This is exactly the same decoding on curves given in \cite[Proposition 2.6]{Ye12}.
\item[(iii)] Let $t\ge 4$. For any real $0<\Gs\le 1$ and $1<d\le \Gs(q-1)/t-1/t$, there exists is a $(q-1,t,d,\Gs(q-1);\F_q/ \F_q)$-codex. It is clear that the expectation of $X$ is  $\mu=\Gd (q-1)$. In Lemma \ref{twise},  put $A=(1-\Gs)(q-1)/2-\Gd (q-1)$, by Lemma \ref{twise} we obtain
\begin{equation}\label{eq:ge10}\Ge=\Pr[X>(1-\Gs )(q-1)/2]\le8\left(\frac{4t\Gd(q-1)+4t^2}{(1-\Gs-2\Gd)^2(q-1)^2 }\right)^{t/2}.\end{equation}
Thus, by Theorem \ref{thm:3.1}, the Reed-Muller code $\RM(q,d,m)$ is an
$(q-1,\Gd,\Ge)$-locally decodable code with $\Ge$ given in \eqref{eq:ge10}.
It is easy to see from  \eqref{eq:ge10} that  $\Ge\le 8\left(\frac{\Gl_{\Gs,\Gd}t}{\sqrt{q}}\right)^t$, where $\Gl_{\Gs,\Gd}=\frac{\sqrt{8}}{1-\Gs-2\Gd}$.
\end{itemize}
}\end{example}
\begin{remark}{\rm For sufficiently large $q$,
by choice of a suitable $t$, local decoding in Example \ref{exm:4.1}(iii) gives much better probability than those in Example \ref{exm:4.1}(i) and (ii).
}\end{remark}

In the rest of this subsection we are going to apply Algorithm 2 and concatenated codex from algebraic geometry codes over $\F_{q^2}$ to get local decoding of Reed-Muller codes. We first consider decoding using codex from the Hermitian function field.

\begin{theorem}\label{thm:4.3}
For any real $0<\Gs,\Gd\le 1$ and integers $4\le t\le q$, $d>1$ satisfying $\Gs<(1-2\Gd)/2$ and $d\le \Gs q$,  the Reed-Muller code $\RM(q,d,m)$ is a $\left(q(q^3-1),\delta,\Ge\right)$-locally correctable code with $\Ge\le 8q\left(\frac{\nu_{\Gs,\Gd}t}{\sqrt{q^3-1}}\right)^{t}$, where $\nu_{\Gs,\Gd}=\frac{\sqrt{8}}{1-2\Gs-2\Gd}$.
\end{theorem}

\begin{Proof} Consider the Hermitian function field over $\F_{q^2}$ defined in Subsection \ref{subsec:curve}. Let $\mQ=\{(0,0)\}$ and let $\mP$ be the set consisting of all ``finite" points except for $(0,0)$. Then for any real $0<\Gs\le 1$ and integers $4\le t\le q$, $d>1$ satisfying $d\le \Gs q$, there exists a $(q^3-1,t,d,\Gs (q^{3}-1),\F_{q^2}/\F_{q^2})$-codex. Applying Algorithm 2 and \eqref{eq:a1}, we conclude that
the Reed-Muller code $\RM(q,d,m)$ is a
$(q(q^{3}-1),\Gd,\Ge)$-locally correctable code with \[\Ge\le 8q\left(\frac{4t\delta (q^3-1) +4t^2}{(1-2\Gs-2\delta)^2(q^3-1)^2}\right)^{t/2}\le8q\left(\frac{\nu_{\Gs,\Gd}t}{\sqrt{q^3-1}}\right)^{t}.\] The desired result follows.
\end{Proof}

Finally, we apply Algorithm 2 and concatenated codex from  the Garcia-Stichtenoth tower.

\begin{theorem}\label{thm:4.4}
Let $q$ be a square prime power and let $e\ge 2$. Fix reals $0<\Gs,\Gd\le 1$. If  integers $4\le t\le  q^e$, $d>1$ satisfy $\Gs<(1-2\Gd)/4$ and $d\le \Gs q$,  then the Reed-Muller code $\RM(q,d,m)$ is an $\left(qn,\delta,\Ge\right)$-locally detectable code with $\Ge\le 8q\left(\frac{4t\delta n +4t^2}{(1-4\Gs-2\delta)^2n^2}\right)^{t/2}$, where $n=q^e(q-1)-1$. Furthermore, the local decoding complexity is $\poly(n,q)$
\end{theorem}
\begin{Proof} Consider the function field $F_e$ in the Garcia-Stichtenoth tower over $\F_{q^2}$ defined in Subsection \ref{subsec:curve}. Then $N(F_e)\ge q^e(q-1)$ and $g(F_e)\le q^e$. Let $\mQ$ be a single ``finite" rational place set and let $\mP$ be the set consisting of other $n=q^{e}({q}-1)-1$ ``finite" rational place. Then  $d(2g(F)+1+t-1)<\rho n<n$ and hence by Proposition \ref{prop:2.6}, there exists an $(n,t,d,\rho n,\F_{q^2}/\F_{q^2})$-codex, where $\rho=3\Gs$.
 Applying the local decoding Algorithm 2 in Subsection \ref{subsec:3.2} gives the desired result.

 Since the codex is constructed from the Garcia-Stichtenoth tower and the code $C_1^{*d}$ is an algebraic geometry code based on this tower, the result on decoding complexity follows.
\end{Proof}

  By taking $t= n/q=q^{e-1}(q-1)$  in Theorem  {\rm \ref{thm:4.4}}, we obtain the results on local decoding of single coordinate.
\begin{corollary}\label{cor:4.4} Let $q$ be a prime power. Let $d>1, t, m$ be positive integers. Let $\Gd,\Gs$ be two reals in $(0,1)$ with $\Gd<\frac{1-4\Gs}2$. Then the Reed-Muller code $\RM(q,d,m)$ with $d\le\Gs \sqrt{q}$  is   $\left(q^2t,\Gd, O\left(\left(\frac{\mu_{\Gd,\Gs}}{\sqrt{q}}\right)^t\right)\right)$-locally correctable, where $\mu_{\Gd,\Gs}=\frac{\sqrt{8}}{1-4\Gs-2\Gd}$ {\rm (}note that $t$ can be arbitrarily large{\rm)}.

\end{corollary}

\subsection{Multiple-point local decoding of Reed-Muller codes}
In this subsection, we analyze local decoding of Reed-Muller codes to recover multiple coordinates simultaneously. Again we apply Reed-Solomon codes, Hermtian codes and algebraic geometry codes based on the Garcia-Stichtenoth tower, respectively. The proofs are almost identical with those in the previous subsection except for replacing $\mQ$ of a single point set by a $k$-point set. We state the results  without proof below.

\begin{theorem}\label{thm:4.5} Let $q$ be a prime power. Let $d>1, t, m, k$ be positive integers. Let $\Gd,\Gs$ be two reals in $(0,1)$ with $\Gd<\frac{1-\Gs}2$.
\begin{itemize}
\item[{\rm (i)}] {\bf (Reed-Solomon code with $t=1$)} If $k+n\le q$ and $d<\frac{\Gs n}k$, then the Reed-Muller code $\RM(q,d,m)$ is an $(k;n,\Gd,\Ge)$-locally decodable code with $\Ge=\frac{2\Gd}{1-\Gs}.$
\item[{\rm (ii)}] {\bf (Reed-Solomon code with $t=2$)} If $k+n\le q$ and $d<\frac{\Gs n}{k+2}$, then the Reed-Muller code $\RM(q,d,m)$ is an $(k;n,\Gd,\Ge)$-locally decodable code with $\Ge=\frac{\Gd-\Gd^2}{(1-\Gs-2\Gd)^2}\times\frac1{n}.$
\item[{\rm (iii)}] {\bf (Reed-Solomon code with $t\ge 4$)} If $k+n\le q$ and $d<\frac{\Gs n}{k+t}$, then the Reed-Muller code $\RM(q,d,m)$ is an $(k;n,\Gd,\Ge)$-locally decodable code with $\Ge=8\left(\frac{4t\Gd n+4t^2}{(1-\Gs-2\Gd)^2 }\right)^{t/2}\times\left(\frac{1}{n}\right)^{t}.$
\item[{\rm (iv)}] {\bf (Hermitian code with $t\ge 4$)}  If $k+n\le q^{3}$ and $d<\Gs q$, then the Reed-Muller code $\RM(q,d,m)$ is an $(k;qn,\Gd,\Ge)$-locally decodable code with $\Ge=8\left(\frac{4t\Gd n+4t^2}{(1-\Gs-\rho-2\Gd)^2 }\right)^{t/2}\times\left(\frac{1}{n}\right)^{t},$ where $\rho=(k+t+q^2-q)/q^2$.
\item[{\rm (v)}] {\bf (GS tower code with $t\ge 4$)} Let  $e\ge 2$. If $t\le  n$ and $k\le  n$, $k+n\le q^{e}({q}-1)$ and $d<\Gs n$, then the Reed-Muller code $\RM(q,d,m)$ is an $(k;qn,\Gd,\Ge)$-locally decodable code with $\Ge=8q\left(\frac{4t\Gd+4t^2}{(1-\Gs-\rho-2\Gd)^2 }\right)^{t/2}\times\left(\frac{1}{n}\right)^{t},$ where $\rho=(2q^{e+1}+qk+qt)/n$. Furthermore, the local decoding complexity is $\poly(n,k,q)$
\end{itemize}
\end{theorem}

Note that we applied Algorithm 1 for the first three local decodings in Theorem \ref{thm:4.5}, while Algorithm 2 is employed  for the last two local decodings in Theorem \ref{thm:4.5}.\\ \\
{\bf Proof of Theorem \ref{thm:2}:} Taking $n\approx \frac{2q}{2q+1}\times q^e(q-1)$ and $k=t=\lfloor {n}/{(2q)}\rfloor$, we obtain Theorem \ref{thm:2} from Theorem \ref{thm:4.5}(v).

\end{document}